\documentclass[final,3p,times]{elsarticle}

\usepackage{graphicx}
\usepackage{lineno}
\usepackage{amsmath}
\usepackage{ulem}
\usepackage{color}

\usepackage{bm}

\usepackage{soul}

\begin{document}
\title{Does a rising tide lift all boats? A wealth exchange model on a dynamic network with economic growth}

\author{Gustavo L. Kohlrausch}
\author{Sebastian Gonçalves}
\affiliation{Instituto de Fisica, Universidade Federal do Rio Grande do Sul,  91501-970 Porto Alegre, RS, Brazil}

\begin{abstract}
Wealth inequality, although an age-old problem, has seen a substantial rise since the early XXI century. The distributions of wealth and income across countries follow a universal pattern, typically manifesting as a two-class division, which suggests that fundamental mechanisms underpin the emergence of these economic disparities. Agent-based models, which allow the rules of interaction between economic agents to be explicitly defined, are particularly well-suited for studying economic systems and analyzing their emergent properties. In this work, we examine a recently proposed dynamic complex network agent-based model \cite{kohlrausch2024wealth} within the context of a growing economy. The model evolves via three alternating processes: independent stochastic wealth growth of each agent, wealth exchanges between connected agents, and the rewiring of connections within the complex network. The wealth growth of each agent is governed by a stochastic process characterized by two parameters: a drift term $\mu$, representing economic growth, and volatility $\sigma$, reflecting heterogeneity in productivity. We analyze the outcomes for various values of a social protection factor $f$, which favors the poorer agent in each transaction. Higher values of $f$ amplify the effect of economic growth: while increasing $\mu$ reduces inequality, increasing $\sigma$ has the opposite effect. In this context, economic growth benefits the poorest agents only when strong social protection is in place.
\end{abstract} 
\maketitle

\section{Introduction}
Economic inequality has become a pressing challenge in modern societies. Indeed, disparities in income, wealth, and access to education and healthcare have widened considerably since the beginning of the 21st century~\cite{Alvaredo2018, CapitalXXI,IncomeChancel,saez2016wealth}. Notwithstanding marked differences in geography, culture, and historical context, income and wealth distributions across countries exhibit a strikingly similar structure~\cite{Chakrabarti2013}. This universality has attracted the attention of economists for over a century, prompting the proposal of numerous distribution functions to fit the observed data~\cite{kleiber2003}. A particularly insightful approach decomposes the income distribution into two distinct regimes~\cite{Yakovenko2001, Ludwig2022}, where the upper tail, which typically accounts for less than $5\%$ of the population, follows a power-law pattern, while the bulk of the distribution conforms to a Gamma distribution~\cite{Clementi2005}. This universal two-regime structure strongly suggests that a set of fundamental mechanisms drive economic inequalities independently of national specificities. Unraveling these underlying mechanisms is therefore essential for devising effective policies to prevent or mitigate such stratification.

In this sense, one simple and convenient way of doing that research is made by the proposal of distinct agent-based models. This approach consists in simulate a virtual society where we can set the rules of interaction between economic agents and observe its influences on the macroscopic quantities. Following this idea, a great variety of agent-based models have been recently studied \cite{Benhur, Nener2021,Iglesias2021,Yakovenko2009,Laguna2021, Chakraborti2000, Chatterjee2004, Li2019,boghosian2022economically}. The main strength of this approach is the ease with which different factors can be incorporated into the system, such as taxation rules \cite{Iglesias2021} and rationality \cite{Laguna2021}. Nevertheless, most of these works adopt a mean-field description, wherein agents interact randomly with all others, and assume a strictly conservative economy, thereby neglecting economic growth. Only a limited number of contributions integrate complex networks into agent-based models \cite{Ma2013, Braunstein2013, kohlrausch2024wealth}, thus introducing topological features into the analysis. Similarly, few studies account for economic growth within the agent-based framework \cite{bouchaud2000wealth,Liu2021}. To the best of our knowledge, however, no existing work has analyzed the simultaneous influence of both network topology and economic growth.

Another central issue in the study of economic inequality concerns to its dynamical properties. Beyond the universal features of income and wealth distributions, a marked rise in top income inequality, characterized as a fattening of the tail of the distribution, has been observed in numerous countries since the final decades of the XX century. This brings a debate about the economical forces that are letting to this dynamics in the top wealth \cite{CapitalXXI, saez2016wealth, gabaix2016dynamics,bricker2016measuring}. However, most agent‑based models concentrate on the characterization of stationary distributions. In the present work, we move beyond this conservative framework by incorporating a stochastic independent growth mechanism into a recently proposed dynamic complex network model \cite{kohlrausch2024wealth}. We thereby investigate the influence of economic growth and wealth exchange on both the stationary distributions and the dynamical characteristics of the system.
We describe the model in Sec.~\ref{secmodel}, the results in Sec. \ref{secresults} and in Sec. \ref{secconclu} we present our conclusions. 

\section{Model}\label{secmodel}
We will present in this section the dynamical evolution of the adopted model across three subsections. First of all, the topological dynamics of the system, from the network construction to its evolution over simulation time. 
Then we will expose the economic trades process and lastly the economic growth of the system. We emphasize that although these three processes are implemented sequentially in the simulation in the same way they are presented here, they are interdependent, and changing their order does not alter any of the results reported in this work. The completion of the three processes, the rewire of connections, the exchange between connected agents and the random independent growth of all agents constitutes one Monte Carlo Step, which is our unity of time.

\subsection{Network creation and dynamics}

The construction and evolution of the network follow the model proposed in Ref.~\cite{gustavo2023wealth}. We begin by assigning each agent a wealth ($\omega$) and a risk factor ($\alpha$), both independent random variables uniformly distributed in $[0,1)$. Next, we select a set of $z$ agents and fully connect them to form the initial network configuration. It is important to note that the parameter $z$ does not affect the stationary or dynamic properties of the model; its only role is to initialize the network.\footnote{Although we tested different values of $z$ in the range $[3, N/2]$, all results presented here use $z=3$.}  

Then, we randomly choose an unconnected agent $j$ and connect it to an existing network agent $i$ with probability  

\begin{equation}\label{prob_con}
 P_{\textrm{connection}}^{i,j} = \frac{\omega_i(t) +\omega_j(t)}{\sum_l \omega_l(t)},
\end{equation}  

where the sum over $l$ runs only over agents that already have at least one connection, and $\omega_i(t)$ denotes the wealth of agent $i$ at time $t$. For each newly added agent $j$, we attempt connections with all currently connected agents, limiting $j$ to at most $z$ connections. This procedure continues until all $N$ agents have been added to the network.  

The rewiring process begins by randomly selecting a pair of agents $(i',j')$. If the pair is not connected, a new connection is created with probability given by Eq.~\ref{prob_con}; if they are already connected, the connection is broken with probability $Q = 1 - P_{\textrm{connection}}^{i',j'}$. To ensure that all agents can participate, we select $N/2$ pairs to rewire their connections in each rewiring step.

\subsection{Economic trades}
In the wealth exchange process, each agent trades wealth with all of its connections according to the Yard-Sale rule~\cite{Benhur, Hayes}. Under this rule, the amount of wealth exchanged between agents $i$ and $j$ is given by
\begin{equation}
    d\omega = \min[\alpha_i \omega_i(t),\; \alpha_j \omega_j(t)].
\end{equation}

To determine the direction of wealth flow, we define the probability that the poorer agent wins the transaction as~\cite{Scafetta}
\begin{equation}\label{prob_exc}
    P_{\textrm{exchange}}^{i,j} = \frac{1}{2} + f \times \frac{|\omega_i(t) - \omega_j(t)|}{\omega_i(t) + \omega_j(t)},
\end{equation}
where $f$ is the social protection factor, ranging between $0$ and $1/2$. After the trade, the wealth of agents $i$ and $j$ becomes
\begin{align*}
 \omega_i(t+1) = \omega_i(t) + d\omega &&  \omega_j(t+1) = \omega_j(t) - d\omega,
 \end{align*}
where agent $i$ is the winner of the transaction. Thus, total wealth is conserved throughout the wealth exchange process. We also emphasize that all connections are considered in the economic trade process, and every exchange between connected agents occurs once per time step. Moreover, the order in which these exchanges are performed does not affect the results.

\subsection{Stochastic economic growth}
In this work we will go beyond the usual wealth conservative scenario employed on the literature \cite{gustavo2023wealth, Benhur, Laguna2021,Yakovenko2009,Chakraborti2000,calvelli2023wealth}, adding a economic growth to the dynamic network model \cite{gustavo2023wealth}.
For this purpose, a simple and commonly employed approach within wealth distribution studies is to consider independent stochastic growth for each agent~\cite{stojkoski2022income, stojkoski2022ergodicity, gabaix2016dynamics, kemp2022statistical, berman2020wealth}. Thus, at each time step, the wealth of agents evolves according to
\begin{equation}\label{eqprod}
 \omega_i(t+1) = \omega_i(t)(1+ \mu + \sigma dB_i) + \Delta\omega_i(t),
\end{equation}
where $\mu$ represents the overall economic growth, benefiting all agents. $dB$ denotes a Brownian motion, so that agents may lose wealth independently of the exchange process if $\sigma > \mu$, where $\sigma$ is the standard deviation in production, representing the heterogeneity in this process. The last term in Eq.~\ref{eqprod} is the total wealth exchanged by agent $i$ at time $t$ during the exchange process. The production process described by the first term on the right-hand side of Eq.~\ref{eqprod} is a purely random growth process, for which no stationary solution exists~\cite{gabaix2016dynamics, berman2020wealth}, and the variance of the $N \times \omega$ distribution grows without bound. Consequently, the exchange process, more specifically, the social protection factor $f$, acts as a stabilizing force for the system, ensuring the existence of a stationary solution.

Since we are now dealing with growing wealth, to optimize the simulation we rescale the wealth of all agents at each time step. This avoids working with very large $\omega$ values, which would occupy unnecessary memory and increase computational cost. Thus, after performing all wealth exchanges on the network and the stochastic growth of all agents, we apply
\begin{equation}\label{eq_rescale}
 \omega_{i} = \omega_{i}\frac{N}{\Omega},
\end{equation}
and therefore the wealth of agents remains confined to the interval $\omega = [0, N]$. We perform simulations for systems with size $N=10^3$ which evolve for $t=4\cdot10^4 MCS$. The results are averaged over $10^3$ independent samples.

\section{Results}\label{secresults}

\begin{figure}
    \centering
    \includegraphics[width=1\linewidth]{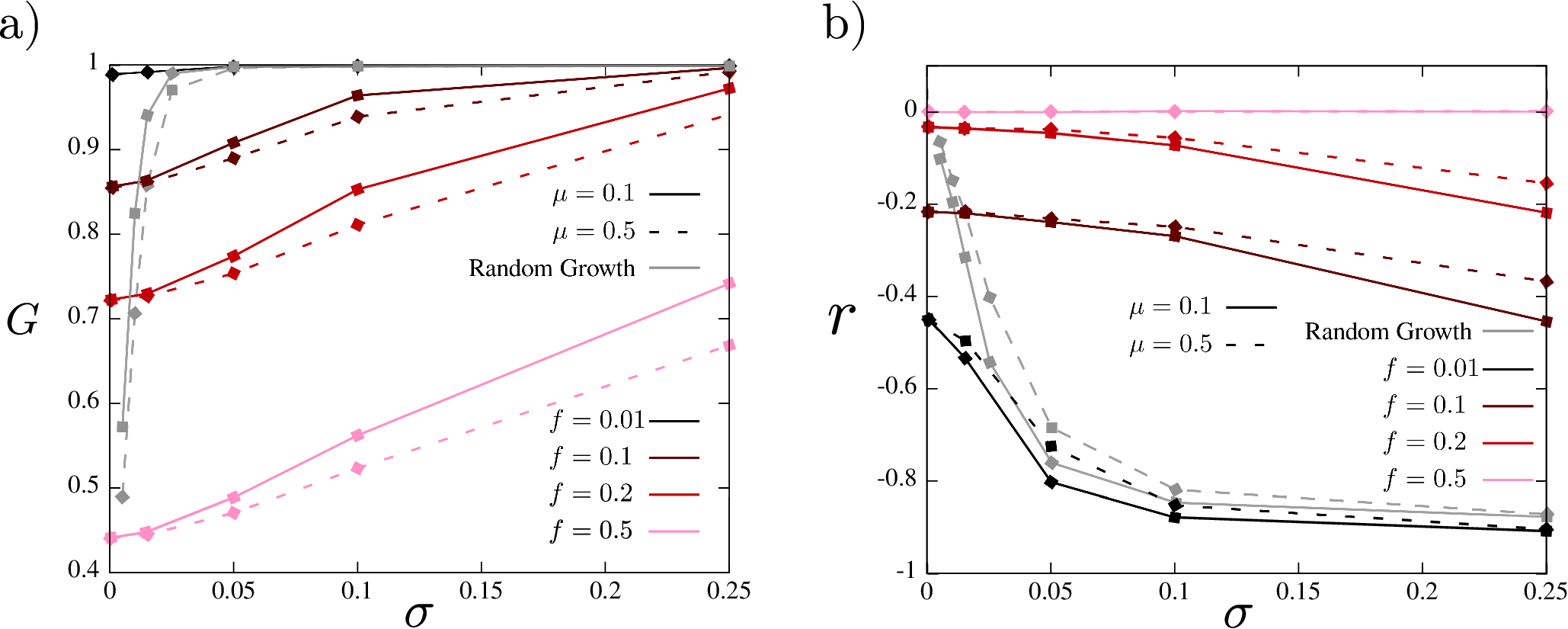}
    \caption{a) Gini index as a function of the standard deviation ($\sigma$) in the production process for different values of social protection ($f$) and economic growth ($\mu$). b) Assortativity of the network as a function of $\sigma$ for different values of $f$ and $\mu$.}
    \label{fig:gini_r}
\end{figure}

We begin the results analysis by looking at the stationary values of the Gini index and assortativity as a function of $\sigma$ (Fig.\ref{fig:gini_r}). We see that increasing the heterogeneity parameter gradually increases the value of $G$ for all values of $f$, thus we have $G=1$ as $\sigma\rightarrow\infty \ \forall f$. However, the parameter $\sigma$ does not seem to have the same asymptotic effect on assortativity. For $f=0.01$ for example, this quantity stabilizes at $r=-0.9$ while for $f=0.5$ we have $r=0$ for all studied values of $\sigma$. The increase in $\sigma$ accentuates disassortativity for $f=0.1$ and $0.2$ across the entire studied spectrum and for $\sigma=[0.015,0.1]$ with $f=0.01$. We note, however, that the connection rule given by Eq.\ref{prob_con} determines that given the condensation of wealth into a single agent, this agent must concentrate all network connections and therefore we have $r=-1$. Thus, we should find a completely disassortative network for $\sigma\rightarrow\infty$, but the network topology is more robust to variations in production heterogeneity compared to economic quantities, especially for higher values of social protection.

Economic growth, represented by increasing $\mu$ from $0.1$ to $0.5$, attenuates inequalities in the system when $f>0.01$, characterized by a decrease in $G$, an effect that is more pronounced for larger values of $\sigma$. The effects of increasing $\mu$ on assortativity are similar, with a more egalitarian system, $r$ approaches zero, which is more evident when the standard deviation of wealth production is larger. However, for $f=0.01$, the values of $r$ for $\mu=0.1$ and $0.5$ become closer as $\sigma$ increases. In this case, where wealth condensation already exists, the increase in $\sigma$ merely dissolves the presence of \textit{hubs} in the network, making it resemble a star-like network.

In order to compare the exchange dynamics with the wealth production dynamics, we also consider purely stochastic growth, thus disregarding the last term in equation \ref{eqprod}, represented by the gray lines in Figure \ref{fig:gini_r}. Since for this case there is no stationary solution, we consider this growth up to $4\times10^4$ MCS, such that the increase in $\sigma$ only accelerates wealth condensation in the system while $\mu$ has the opposite effect. We see that in this case, the values and behavior of $r$ are very close to those with $f=0.01$, only when $\sigma\rightarrow0$ do the two models diverge, where the model with wealth exchanges is more disassortative. Although the Gini index for purely random growth is much lower than that for $f=0.01$ when the production standard deviation is low, it presents wealth condensation for $\sigma>0.015$.

\begin{figure}[!h]
 \centering
 \includegraphics[width=1\columnwidth]{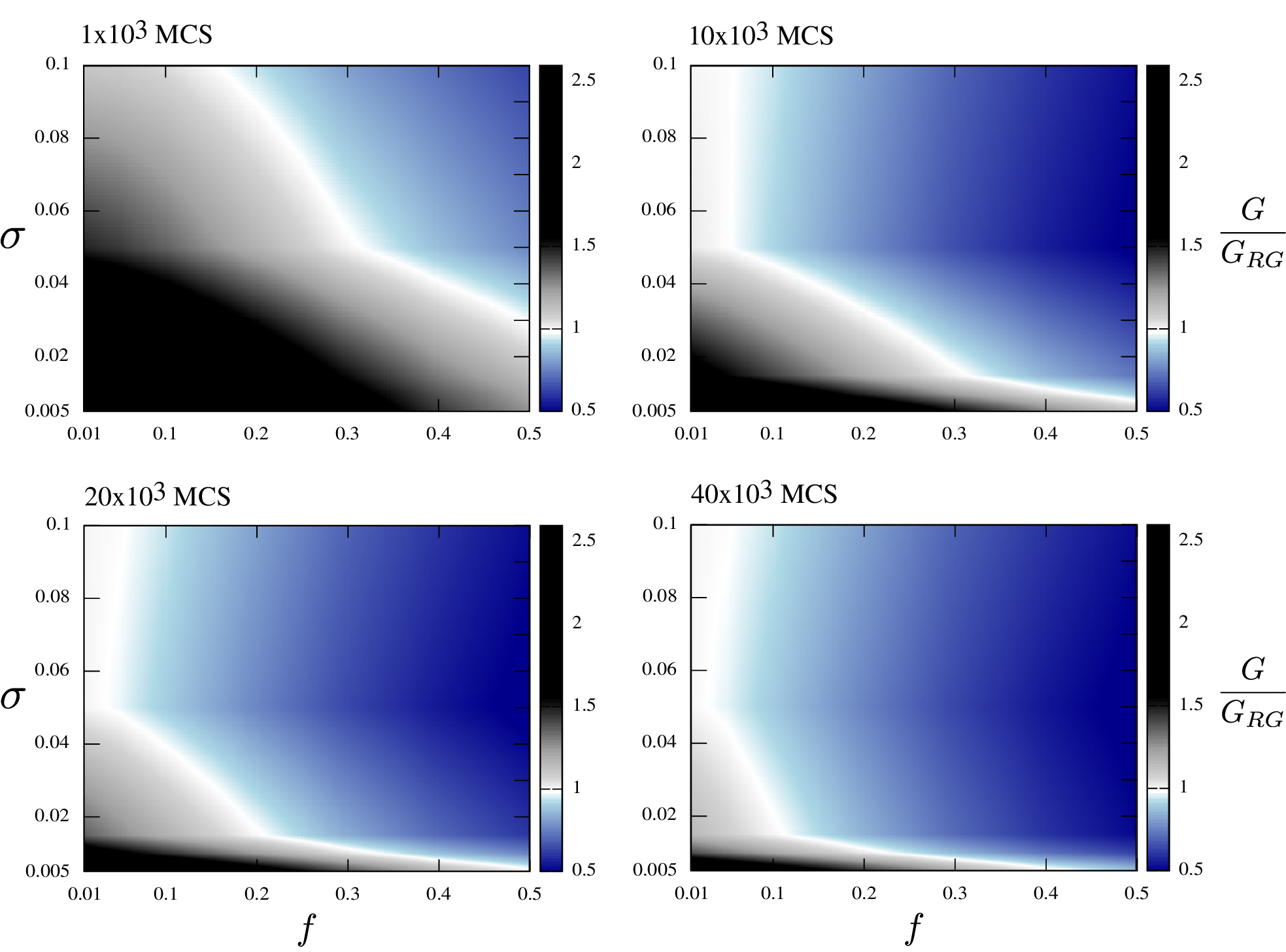}
    \caption{Ratio between the Gini index measured in the model with wealth exchange and stochastic growth ($G$) and in the model without exchanges ($G_{RG}$) as a function of $f$ and $\sigma$ for different simulation times with $\mu=0.5$.}
    \label{fig_dif_varios}
\end{figure}

To better understand the dynamic effects of both wealth production and exchange, we compare the temporal evolution of the macroscopic quantities $G$ and $r$ with those found in purely random growth ($G_{RG}$ and $r_{RG}$). We start by looking at the ratio of the Gini indices of the two models at four different times. Figure \ref{fig_dif_varios} shows the quantity $G/G_{RG}$ near the beginning of the simulation ($1\times10^3$ MCS), at two intermediate times ($10$ and $20\times10^3$ MCS), and at the end ($40\times10^3$ MCS).

When the standard deviation in production is very small, $\sigma=0.005$, purely random growth has a very slow dynamics and $G/G_{RG}>1\ \forall f$ for $t\leq20\times10^3$ MCS, and the system with wealth exchange becomes more egalitarian only at the final simulation times for $f\geq0.4$.
At low standard deviation values, $\sigma\leq0.05$, we observe that the exchange system is already more egalitarian at the initial simulation times when social protection is sufficiently high. As the simulation advances, the model with only wealth production tends toward condensation, and the blue region in Figure \ref{fig_dif_varios} (where the exchange model has a lower Gini) grows.

At $t=1\times10^3$ MCS we find a line $G/G_{RG}=1$ (white in Figure \ref{fig_dif_varios}) with two distinct slopes, while at $t=10\times10^3$ MCS we find three distinct slopes. For the initial simulation time, the slope change occurs at $\sigma=0.05$, whereas at the intermediate time the new change occurs at $\sigma=0.015$, which are the points where $G_{RG}=1$ and $G_{RG}\approx1$, respectively, when $t=40\times10^3$ MCS. At high standard deviation values, $\sigma>0.05$, the exchange system is always as egalitarian as or more egalitarian than purely random growth for $t>10\times10^3$ MCS, where as the simulation advances the line representing $G/G_{RG}=1$ tilts to the right as the system without wealth exchange reaches higher Gini values. However, at the initial simulation times, economic inequality is higher in the population when agents carry out financial transactions for $f\leq0.3$ even with high values of $\sigma$. Thus, wealth production affects the system dynamics in a non-trivial way, and we can observe a more egalitarian distribution in the model without exchanges at early times, but as the simulation proceeds this situation reverses when $\sigma$ and $f$ are sufficiently high.

\begin{figure}[!h]
 \centering
 \includegraphics[width=1\columnwidth]{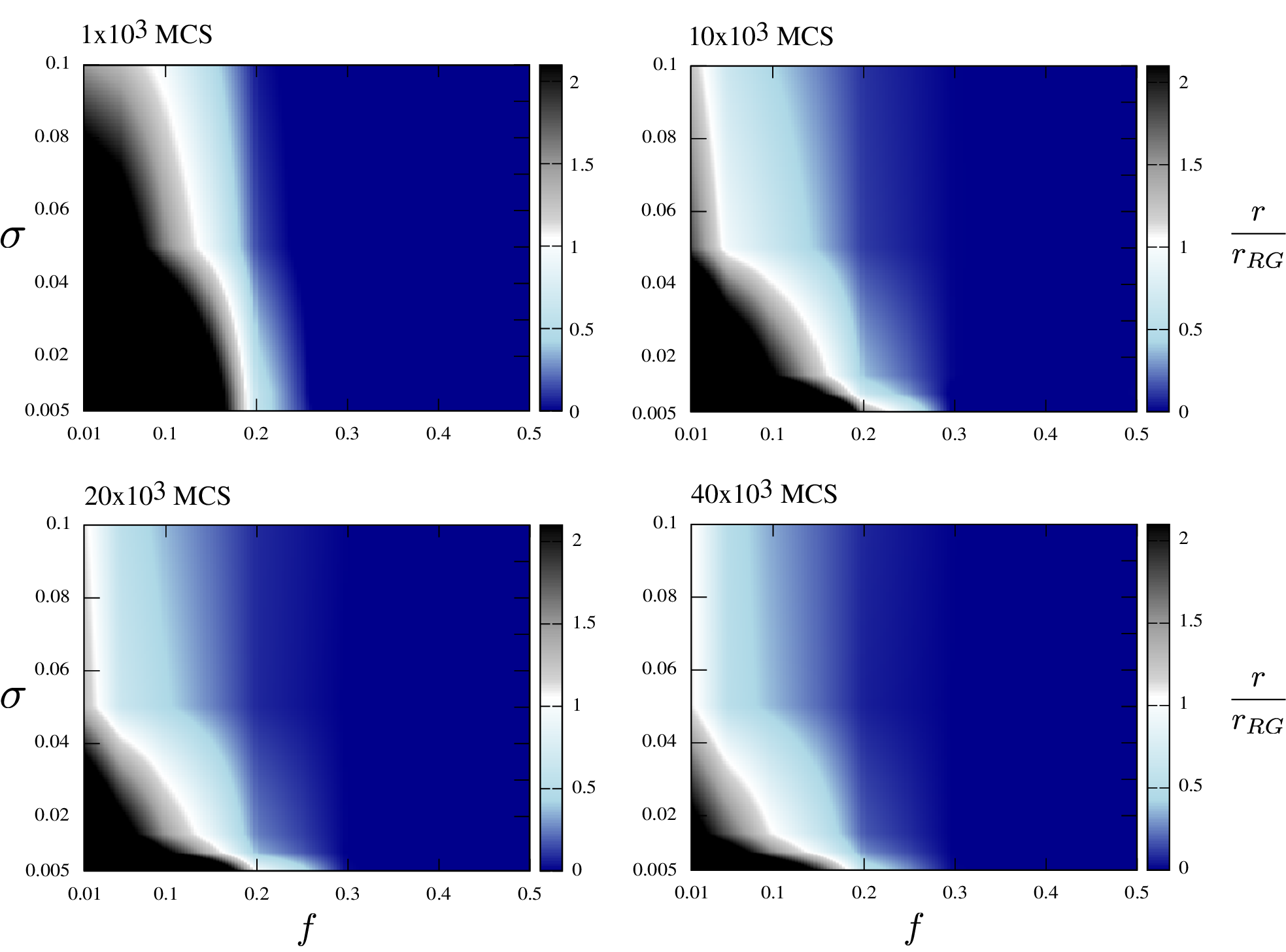}
    \caption{Ratio between the assortativity measured in the model with wealth exchange and stochastic growth ($r$) and in the model without exchanges ($r_{RG}$) as a function of $f$ and $\sigma$ for different simulation times with $\mu=0.5$.}
    \label{fig_dif_assorc}
\end{figure}

Let us now look at the effects of wealth production on the topological evolution of the system. In Figure \ref{fig_dif_assorc} we present the ratio between the assortativity of the model with and without exchanges ($r/r_{RG}$) for the same four simulation times discussed in Figure \ref{fig_dif_varios}. In this figure, a region with $r/r_{RG}\approx0$ (dark blue) is already present at $t=1\times10^3$ MCS for $f>0.2$ for all studied values of $\sigma$, which arises from the low assortativity values in the exchange model under strong social protection, as seen in Figure \ref{fig:gini_r} b). As we advance the simulation, this region remains almost unchanged, because in the presence of high $f$, economic exchanges quickly stabilize the system topology. However, when social protection is weak ($f=0.01$) we have $r/r_{RG}\geq1$ for any considered time, only at the end of the simulation does purely random growth reach assortativity values close to those of the exchange model for $\sigma\geq0.05$, as shown in Figure \ref{fig:gini_r} b).

In the intermediate region of $f$ between $0.01$ and $0.2$, we observe that the ratio between assortativities $r/r_{RG}$ decreases as the simulation progresses. For $f=0.1$, for example, the exchange model is more disassortative at the beginning of the simulation for $\sigma<0.1$, whereas by the end we have $r/r_{RG}=1$ at $\sigma=0.015$. Since for this value of social protection the exchange model stabilizes quickly, the decrease in the ratio $r/r_{RG}$ comes from a more disassortative behavior of the random growth model. Nevertheless, a region with $r/r_{RG}>1$ is still observed for $f<0.2$ when the standard  deviation of wealth production is sufficiently low. Thus, exchanges between agents stabilize the network much faster than wealth production. Even at low values of social protection, where the exchange model exhibits slow dynamics, purely random growth showed much later effects. However, the influence of wealth production on system dynamics is more significant at intermediate values of $f$ and $\sigma$. For both high and very low values of $f$, wealth exchanges accelerate the system dynamics compared to the stochastic production model. However, in the case $f=0.01$, transactions between agents accelerate wealth condensation, while for $f>0.2$ they lead the system to a more egalitarian distribution. For values of $\sigma\approx0$, the model without wealth exchanges exhibits very slow dynamics, showing an influence of production only at very long times.

\begin{figure}
    \centering
    \includegraphics[width=1\linewidth]{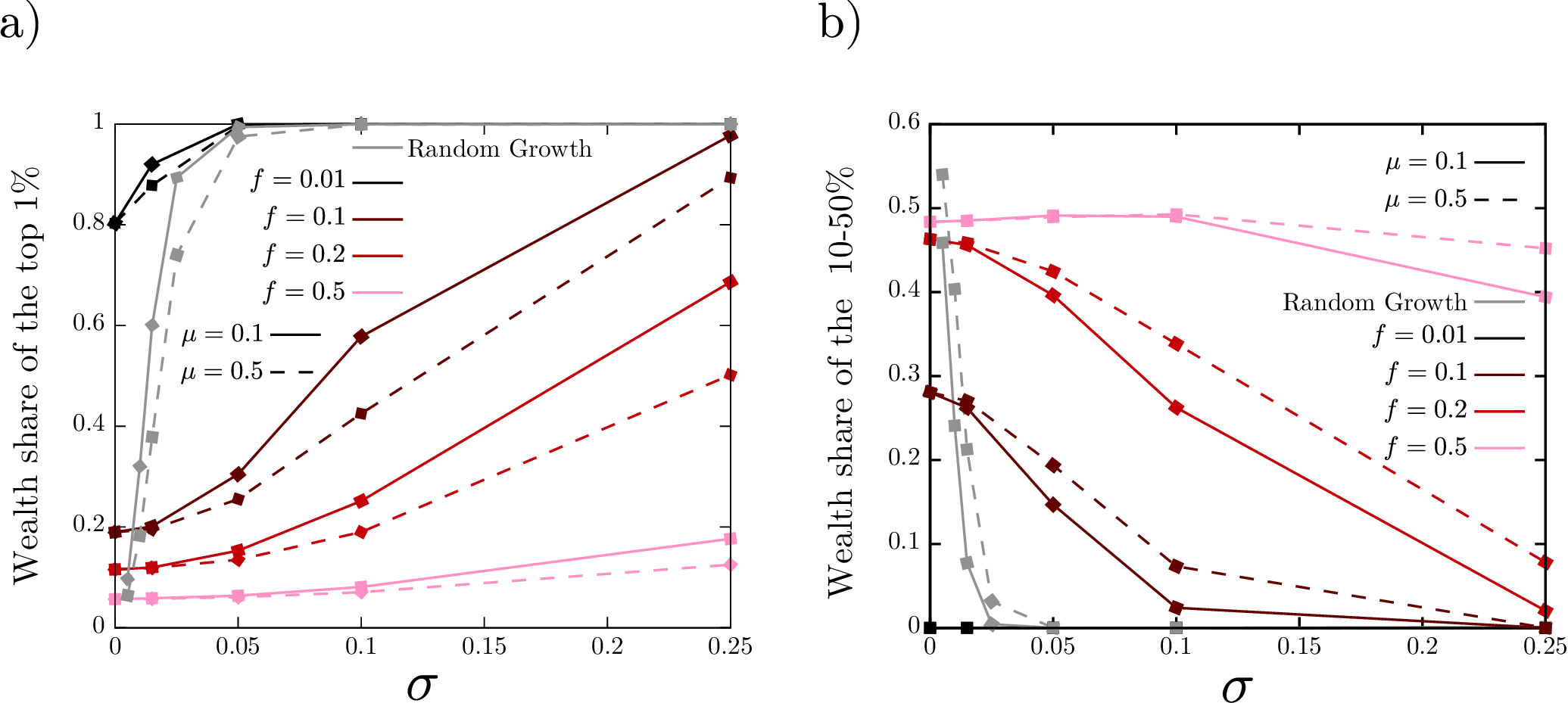}
    \caption{Share of total wealth held by the a) richest $1\%$ and b) $10-50\%$ of the system as a function of the standard deviation in production for different values of $f$ and $\mu$. In gray, we also consider a system without wealth exchanges, i.e., purely stochastic growth, after $4\times10^4$ MCS.}
    \label{fig:share}
\end{figure}

Looking at the wealth distribution in the top percentile (Fig.\ref{fig:share} a)) we see that the increase in inequality due to production heterogeneity is especially caused by the accumulation of wealth in the richest $1\%$ of the population. For $f=0.1$, for example, the top percentile of the distribution holds approximately $20\%$ of total wealth when $\sigma=0$, which grows rapidly to over $95\%$ for $\sigma=0.25$ and $\mu=0.1$. For $f=0.01$ the system condenses all wealth into the richest $1\%$ even for very small values of production heterogeneity, $\sigma\geq0.05$. Furthermore, we see that the model without exchanges is quite susceptible to small variations in $\sigma$, noted by the abrupt growth in the wealth of the top percentile for small changes in $\sigma$ when this parameter is close to zero.

The growth of the economy reduces the wealth of the top percentile, especially for larger values of $\sigma$, in agreement with the Gini index results.  Looking at the wealth of the middle class (Fig.\ref{fig:share} b)), we see that economic growth benefits this strata of the population only when social protection is sufficiently strong. However, for $f=0.5$ the increase in $\mu$ has no effect on the middle $40\%$ of the distribution for $\sigma\leq0.1$, where the wealth of this group remains practically constant even with a higher variance in production. For smaller values of $f$, economic growth has little effect compared to the loss of wealth caused by the growth of $\sigma$, where the middle class shows little to no wealth for sufficiently high values of $\sigma$.

\begin{figure}
 \centering
 \includegraphics[width=1\columnwidth]{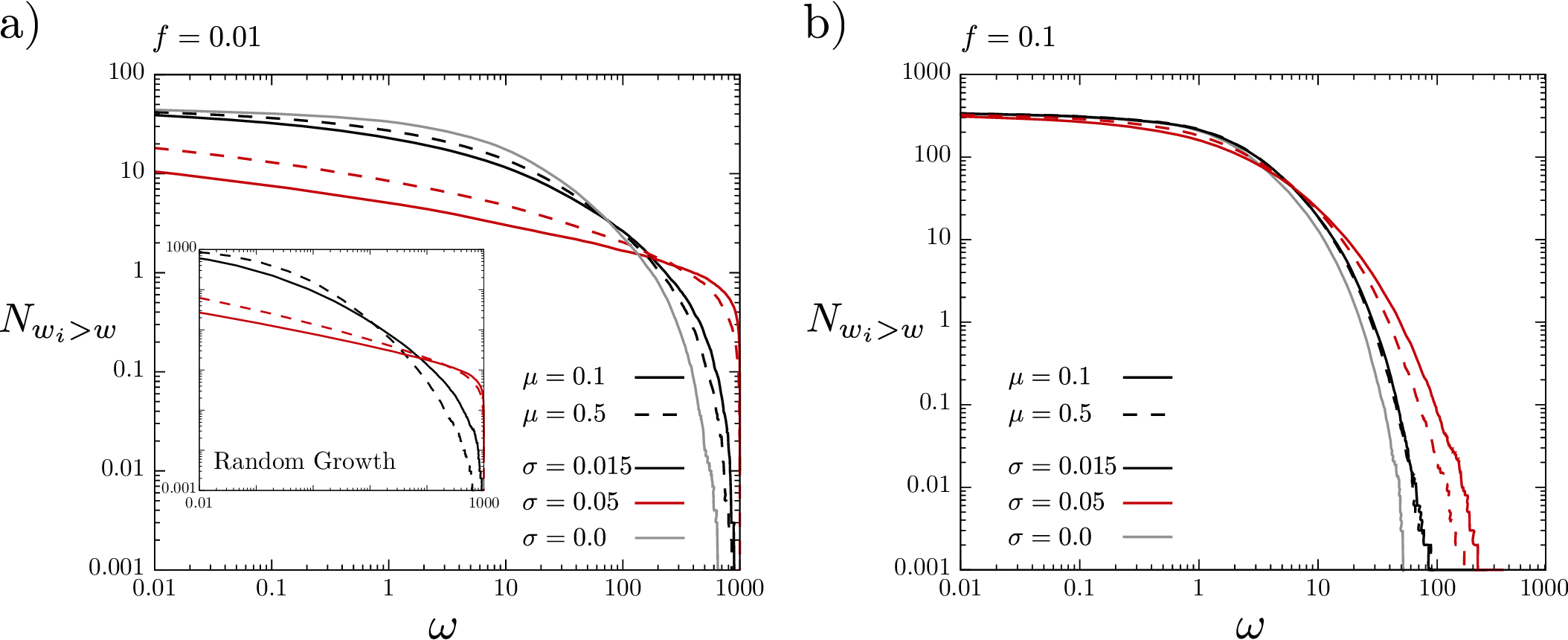}
    \caption{Cumulative wealth distribution for a) $f=0.01$ and b) $f=0.1$ with different values of $\sigma$ and $\mu$. Inset: $N_{\omega_i>\omega} \times \omega$ for the purely random growth model using the same values of $\sigma$ and $\mu$.}
    \label{fig_agent_wealth_growth}
\end{figure}

In Figure \ref{fig_agent_wealth_growth} we present the cumulative wealth distribution for $f=0.01$ and $0.1$ for different values of $\mu$ and $\sigma$. For comparison, we also show the cumulative distribution for a model without wealth production, obtained by taking the limit $\sigma=0$, and for the model without wealth exchanges (inset). As seen earlier, the increase in production heterogeneity accentuates inequalities in the system, which is expressed here by the rightward shift of the curve $N_{\omega_i>\omega}\times \omega$. For $f=0.01$ (Figure \ref{fig_agent_wealth_growth} a)) we see that a small increase in $\sigma$ significantly changes the wealth distribution. In the cumulative distribution we see that precisely only $1\%$ of the agents in the system possess any wealth with $f=0.01$, $\sigma=0.05$ and $\mu=0.1$, characterized by $N_{\omega_i>0.01}=10$, which rises to just over $1\%$ with $\mu=0.5$. The increase in production variance accentuates the maximum wealth attained by an agent in the simulation, which can be seen at the point $N_{\omega_i>\omega}=0.001$, where $1/\text{number of samples}=0.001$, going from approximately $650$ at $\sigma=0$, to $900$ with $\sigma=0.015$ and $999$ at $\sigma=0.05$.

When we increase social protection to $f=0.1$ (Figure \ref{fig_agent_wealth_growth} b)) we see that the increase in  wealth production standard deviation mainly affects the tail of the cumulative wealth distribution. In this scenario, the beginning of the distribution changes very little: we find $N_{\omega_i>0.01} \approx 339$ for $\sigma = 0$ and $N_{\omega_i>0.01} \approx 325$ with $\sigma=0.05$. Thus, the decrease in middle-class wealth shown in Figure \ref{fig:share} b) does not occur due to the bankruptcy of the poorest agents in that group, but rather due to the much faster growth of upper-class wealth, manifested by the smaller slope of the curve $N_{\omega_i>\omega}\times\omega$. In the model without exchanges (inset of Figure \ref{fig_agent_wealth_growth}), even at low production heterogeneity values, $\sigma=0.015$, we see that the cumulative distribution drops rapidly even for small amounts of wealth. Thus, social protection is able to stabilize the cumulative wealth distribution for low values of $\omega$. For $\sigma=0.05$ the cumulative wealth distributions of the model with and without exchanges become similar when $f=0.01$ due to the strong concentration of wealth in the richest agent of the system observed in these two cases.

\begin{figure}
 \centering
 \includegraphics[width=1\columnwidth]{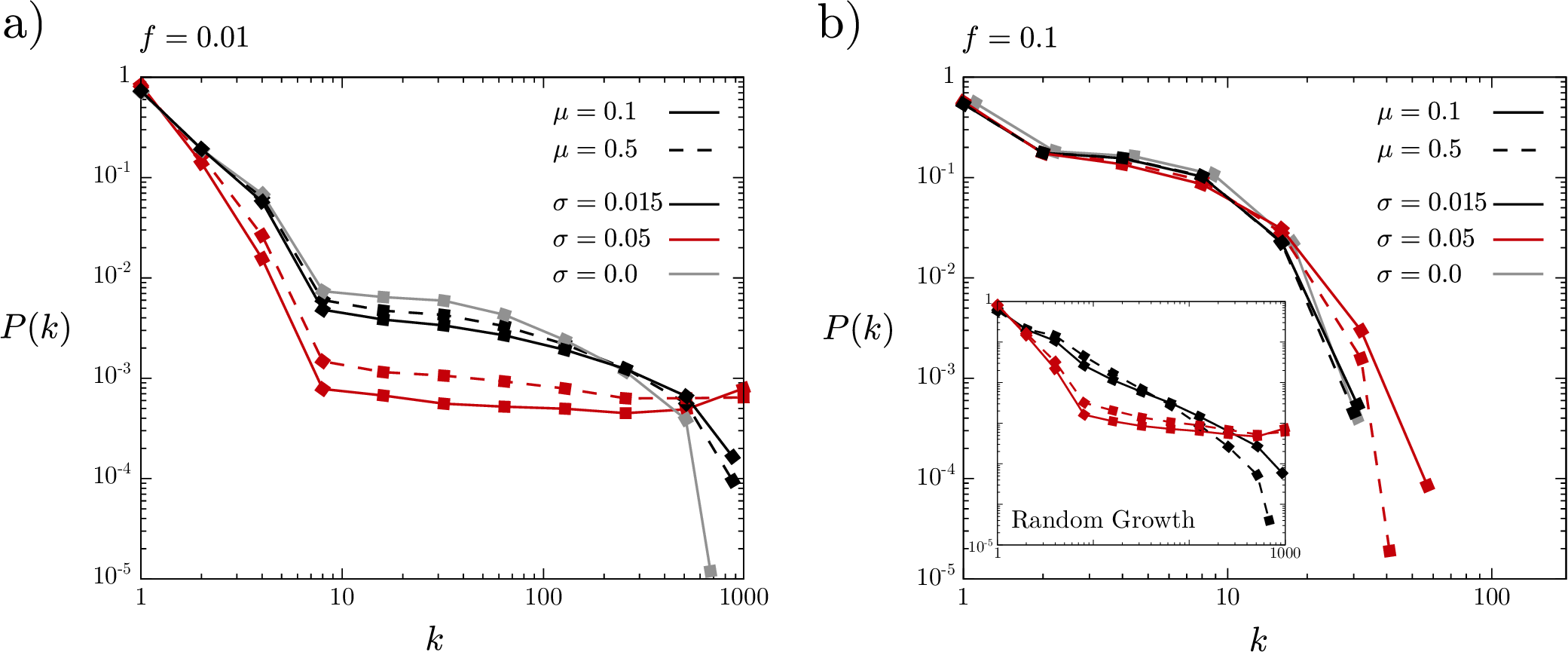}
    \caption{Degree distribution for a) $f=0.01$ and b) $f=0.1$ with different values of $\sigma$ and $\mu$.  Inset: $P(k)\times k$ for the purely random growth model using the same values of $\sigma$ and $\mu$.}
    \label{fig_agent_degree_growth}
\end{figure}

Looking at the degree distribution for $f=0.01$ and $0.1$, Figure \ref{fig_agent_degree_growth} a) and b) respectively, we see that the increase in $\sigma$ affects $P(k)$ in an apparently distinct way for the two social protection values studied. For $f=0.01$, the increase in production standard deviation reduces the probability of finding agents in the range $k=[2,256]$, consequently raising $P(k=1)$ and $P(k>256)$. Thus, the presence of more heterogeneous wealth production concentrates connections among a smaller number of agents, dissolving the hubs present in the network when $\sigma\rightarrow0$, making the network more disassortative. However, the expansion of economic growth increases the probability of finding agents in the range $k=[2,256]$, reducing the values of $|r|$.

When we increase social protection to $f=0.1$, we see that the degree distribution is quite resilient to increases in the parameters $\mu$ and $\sigma$ of wealth production, explaining the small variation of assortativity for higher social protection values shown in Figure \ref{fig:gini_r} b). Nevertheless, we observe the growth of $P(k>16)$, especially for $\mu=0.1$. Thus, even in the presence of considerable favoritism toward poorer agents in wealth exchanges, a more heterogeneous production process accelerates the growth of the rich relative to the rest of the system, favoring them in network formation. In the model without wealth exchanges (inset of Figure \ref{fig_agent_degree_growth}), we again see a similarity between its results for $\sigma=0.05$ and those of the model with exchanges when $f=0.01$, arising from wealth condensation in both models. For $\sigma=0.015$, the model without exchanges exhibits a smoother decay of $P(k)$ compared to the results for $f=0.01$, also evidenced by the less disassortative character of the former, shown in Figure \ref{fig:gini_r} b). Therefore, wealth exchanges favor the emergence of an upper class of strongly connected agents only at lower values of $\sigma$ and $f$, with $f\neq0$, whereas at higher values of $\sigma$ wealth production leads to condensation.

\begin{figure}[!h]
    \centering
    \includegraphics[width=1\linewidth]{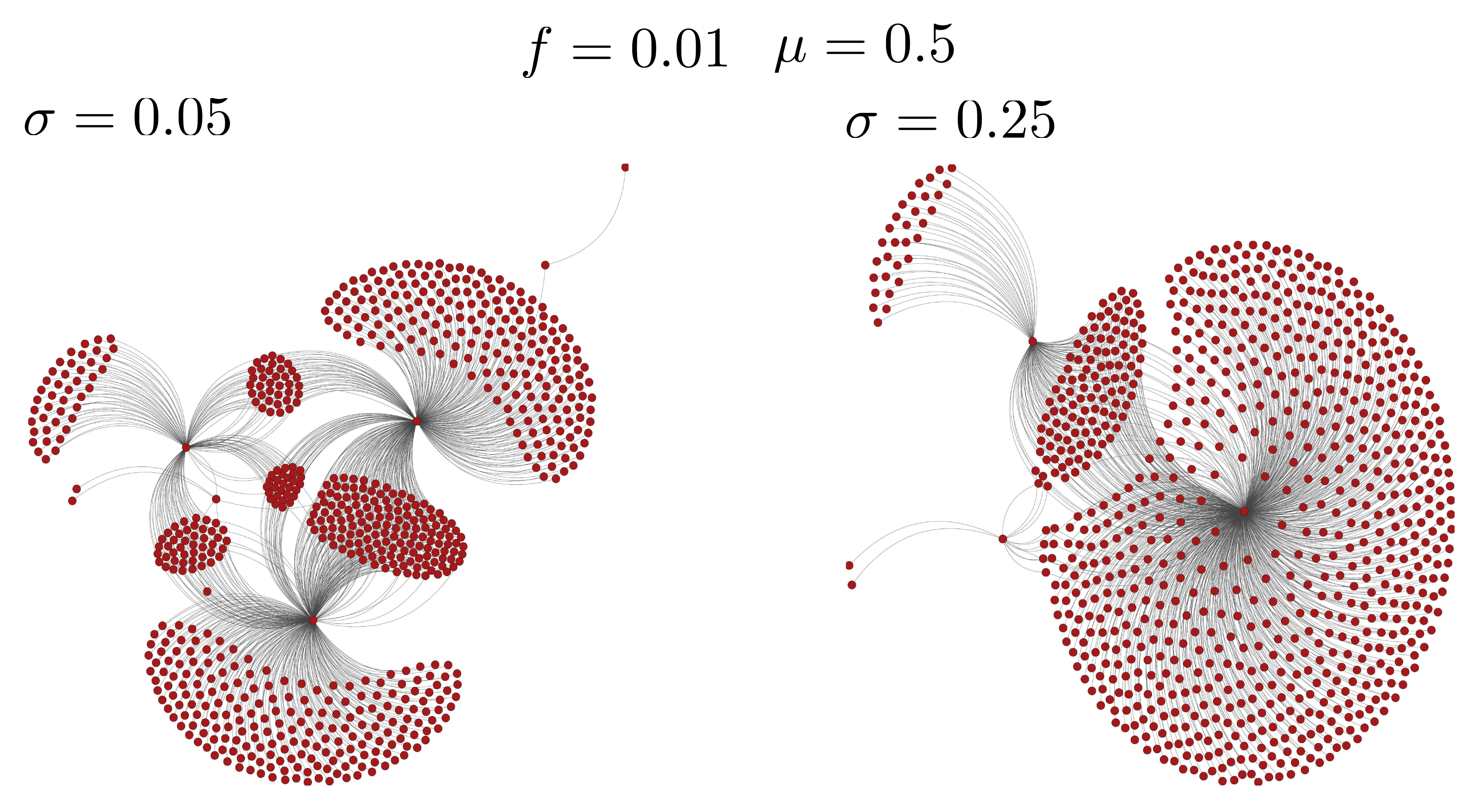}
    \caption{Snapshot of the equilibrium network configuration for $f=0.01$, $\mu = 0.5$ and $\sigma = 0.05$ (left) and $0.25$(right).}
    \label{fig:rede_f=0.01}
\end{figure}

To better illustrate the dissolution of \textit{hubs} in the network by the increasing of $\sigma$ in the $f=0.01$ 
 case, we present in Figure ~\ref{fig:rede_f=0.01} snapshots of the equilibrium network for two distinct values of $\sigma$. It is clear from this example and from the degree distribution that the increase in the production variance leads the network to a star-like structure, concentrating the connections in a single agent.
 
\section{Conclusions}\label{secconclu}
In this work we went beyond the conservative wealth approximation by adding to the model a stochastic growth process independent for each agent. This process is characterized by two distinct parameters: the economic growth $\mu$, which appears as an additive term to all constituents of the system, and the production standard deviation $\sigma$, associated with heterogeneity in wealth production processes in society. The increase in $\mu$, although reducing inequality in the system, has much smaller effects compared to those caused by increases in $\sigma$, in agreement with the results in the literature \cite{kemp2022statistical}, and $f$. Analyzing the stationary quantities and the dynamic characteristics of the system, we notice a non‑trivial relationship between the effects of $f$ and $\sigma$. For values of $f$ close to zero, the exchange of wealth among agents accelerates condensation in the top percentile of the distribution compared to the model without exchanges. Although for low values of $f$ we observe slower dynamics in the system topology, the model with exchanges evolves more rapidly than purely random growth, even for high values of $\sigma$, which accelerates the concentration of connections and wealth in both models. For $f>0.2$, we see that the exchange model stabilizes the network topology already at the beginning of the simulation, and increasing $\sigma$ makes the network less disassortative, especially at $f=0.5$, where the network is non‑assortative for all studied values of $\sigma$. In this region of $f$, we observe a lower Gini index in the model without exchanges in the initial moments of the simulation when $\sigma\rightarrow0$, which reverses by the end of the simulation. Thus, exchanges among agents stabilize the wealth distribution, particularly in the middle and lower classes, while the production process favors the growth of agents in the upper classes, leading to higher values of $G$ as $\sigma$ increases.

For intermediate values of $f$, the evolution of the model exhibits more complex dynamics, showing a greater dependence of variance on the wealth production process. In general, in the region $0.01<f<0.2$, we observe the model with wealth exchanges to be more economically unequal and disassortative in the initial moments of the simulation. This situation persists at the end of the simulation for low values of $\sigma$, but reverses when heterogeneity is sufficiently strong, which depends on the value of $f$. Thus, the wealth production process has a much more pronounced influence on the dynamics of wealth inequality for intermediate values of social protection. In this region, $f$ is able to favor agents mainly from the middle class, avoiding the rapid wealth condensation that occurs at low values of $f$, while agents from the lower class still go bankrupt. It is the presence of this middle class with considerable wealth that is able to maintain a significant wealth production dynamics in the exchange model. However, the poorer half of the population holds a significant amount of wealth only under very strong social protections, for which wealth exchanges rapidly stabilize the system. Therefore, the inclusion of stochastic growth into the dynamic network model revealed a very interesting phenomenology, both in the stationary quantities and in the system dynamics, especially in the interval $f=[0.01,0.2]$, so that the temporal relationship between exchanges and wealth production is quite complex and sensitive to the model parameters.
\section*{Acknowledgments}
The authors acknowledge support from Brazilian funding agency Conselho Nacional de Desenvolvimento Científico e Tecnológico (CNPq), GK for PhD scholarship and SG for grant \#314738/2021-5.

\bibliography{references.bib}

\end{document}